
\input phyzzx

\input tables
\def\dps{\Delta_\psi}
\def\dph{\Delta_\phi}
\def\nps{n_\psi}
\def\nph{n_\phi}
\def\mps{m_\psi}
\def\mph{m_\phi}
\def\myfoot#1{\attach#1\vfootnote{#1}}
\def\pf#1{{\it Phys. of Fluids} {\bf #1}}
\def\jfm#1{{\it J. Fluid Mech.} {\bf #1}}
\def\sam#1{{\it Stud. Appl. Math.} {\bf #1}}
\def\pha#1{{\it Physica} {\bf #1}}

\def\prr#1{{\it Phys. Rev.} {\bf #1}}
\def\np#1{{\it Nucl. Phys.} {\bf B#1}}

\def\jmath#1{{\it J. Math. Phys.} {\bf #1}}

\REF\Kraa{R.H. Kraichnan, \pf {10}, 1417 (1967).}
\REF\Krab{R.H. Kraichnan, \jfm {47}, 525 (1971).}
\REF\Saff{P.G. Saffman, \sam {50}, 277 (1971).}
\REF\Moff{H.K. Moffatt, in {\it Advances in turbulence}, G. Comte-Bellot
and J.Mathieu, eds, p.284 (Springer-Verlag 1986).}
\REF\Poly{A. Polyakov, ``Conformal Turbulence", Preprint PUPT-1341,
 Bulletin Board: hep-th@xxx.lanl.gov - 9209046.}
\REF\BPZ{A. Belavin, A. Polyakov, and A. Zamolodchikov, \np
{241}, 33 (1984).}
\REF\Lesi{M. Lesieur, {\it Turbulence in Fluids}, (Kluwer, London 1990).}
\REF\Krac{R. Kraichnan, \pf {8}, 575 (1965).}
\REF\Beli{V. Belinicher, V.Lvov, Sov. Phys. JETP {\bf66}, 303 (1987);
V.Lvov, G.Falkovich, \prr {A46} (1992).}
\REF\WaTu{V. Zakharov, V. Lvov, G. Falkovich, {\it Kolmogorov Spectra
of Turbulence} v.1 Wave Turbulence, (Springer Verlag, Heidelberg 1992).}
\REF\DNPZ{V. Djachenko, A. Newell, A. Pushkarev, V. Zakharov, \pha
{D57}, 96 (1992).}
\REF\Krad{R. Kraichnan, \jfm {67}, 155 (1975).}
\REF\Benz{R.Benzi, G.Paladin, A.Vulpiani, \prr {A42}, 3654 (1990).}

\Pubnum={WIS-92/88/Nov-PH}
\titlepage
\date={Nov 1992}
\title{\bf SPECTRA OF CONFORMAL TURBULENCE}
\author{\bf Gregory Falkovich\myfoot{a}{\rm e-mail: FNFAL@WEIZMANN.BITNET}
 and Amihay Hanany\myfoot{b}{\rm e-mail: FTAMI@WEIZMANN.BITNET}}
\address{Department of Physics \break
         Weizmann Institute of Science \break
         Rehovot 76100, Israel}
\vfill
\abstract{ A set of different conformal solutions corresponding to a
constant flux of squared vorticity is considered. Requiring constant
fluxes of all inviscid vorticity invariants (higher powers of the
vorticity), we come to the conclusion that the general turbulence
spectrum should be given by Kraichnan's expression $E(k)\propto
k\sp{-3}$. This spectrum, in particular, can be obtained as a limit of
some subsequences of the conformal solutions. }
\overfullrule=0pt

The problem of small-scale spectrum of two-dimensional turbulence is a
peculiar problem among the variety of turbulent systems. The point is
that dimensional considerations do not give a steady spectrum that
corresponds to the enstrophy (squared vorticity) cascade. The spectrum
obtained from dimensional analysis is $E(k)\propto k\sp{-3}$
\lbrack1\rbrack\ which yields a logarithmic infrared divergence after
substitution into the equations for the correlation functions.
Since any cascade picture assumes turbulence locality, the divergence
makes this spectrum rather suspicious.
Kraichnan's attempt to save the spectrum from nonlocality
by introducing the slow factor $\ln\sp{-1/3}k$ attains convergence only
in the first order of perturbation theory \lbrack2\rbrack\
while the next orders reveal
divergencies with higher powers of the logarithm: $\ln\sp2$ etc. The fact
that the powers of the logarithm increase with the order of perturbation
theory suggests that a substantial renormalization of the index occurs.
No successful attempts to work out the divergencies or to show that
they are cancelled are known to us. The existence of
alternative predictions for the steady spectrum $E(k)\propto k\sp{-4}$
by Saffman \lbrack3\rbrack\ and $E(k)\propto k\sp{-11/3}$ by Moffatt
\lbrack4\rbrack, show that this is still an open problem.

A fairly new approach to the problem has recently been
introduced by Polyakov \lbrack5\rbrack. He suggested to borrow a set of
correlation functions from conformal field theory to satisfy
a chain of equations following from Euler's equation.
Note that the conformal invariance of turbulence should be
considered as a pure conjecture.
It is convenient to start from the Navier-Stokes equation written
for the vorticity field $\omega({\bf x},t)$
$${\partial\omega\over\partial t}+e_{\alpha\beta}{\partial\psi\over
\partial x_\alpha}{\partial\Delta\psi\over\partial x_\beta}=\nu\Delta
\omega\ .\eqn\Euler$$
Here $\psi$ is a stream function giving the velocity field as follows:
$v_\alpha=e_{\alpha\beta}\partial_\beta\psi$. Here and below we use
a shorthand notation $\partial/\partial x_\alpha=\partial_\alpha$.

Our aim is to find a stationary set of equal time correlation functions
$$I_n({\bf x}_1,\ldots,{\bf x}_n)=\langle
\omega({\bf x}_1,t)\ldots \omega({\bf x}_n,t)\rangle\ .$$
The brackets denote an average with some time independent probability
distribution:
$$\sum_{p=1}\sp n\left\langle\omega({\bf x}_1,t)\ldots{\partial\omega(
{\bf x}_p,t)\over\partial t}\ldots\omega({\bf x}_n,t)\right
\rangle=0\ .\eqn\Hopf$$
Such a stationary set is expected to exist in the inertial interval of
scales, i.e., for distances that are much less than the scale of an
external pump (or the scale of an initial distribution for free decay)
and much larger than the viscous scale $a$. It is possible, then, to
neglect viscosity in (1) using instead a careful point splitting
procedure \lbrack5\rbrack\ for the nonlinear term
$$
e_{\alpha\beta}{\partial\psi({\bf x})\over
\partial x_\alpha}{\partial\Delta\psi({\bf x})\over\partial x_\beta}=
\lim_{a\rightarrow0} e_{\alpha\beta}{\partial\psi({\bf x}+{\bf a}/2)
\over \partial x_\alpha}{\partial\Delta\psi({\bf x}-{\bf a}/2)
\over\partial x_\beta}\ ,\eqn\psipsi$$
where ``lim'' implies angle averaging. To calculate different-point pair
correlators like $\psi({\bf x}+{\bf a})\psi({\bf x}-{\bf a})$,
the fusion rule of the type
$$\lbrack\psi\rbrack\,\lbrack\psi\rbrack=\lbrack\phi\rbrack+\ldots
\eqn\ope$$
should be used.
Here we follow the notations of Ref.5 so that $\lbrack\psi
\rbrack$ means the conformal class of $\psi$, i.e. itself together
with the operators $L_{-n_1}\ldots L_{-n_k}\psi$, $L_{-n}$ being
Virasoro generators \lbrack6\rbrack. Both $\psi$ and $\phi$ are presumed
to be taken from a set (primary fields) of some conformal field theory
(in this paper we will consider the so called minimal models
\lbrack6\rbrack). The primary field $\phi$ provides
the main contribution in the operator product expansion (OPE)
(4) in the small-scale region, i.e. it has the smallest conformal
dimension. The important thing is that the scaling indices (dimensions)
of the fields are not additive so generally $\dph\not=2\dps$.
The energy density in the wave number space is expressed via $\vert\psi
\vert\sp2$ and is
$$E(k)=2\pi k\epsilon(k)
\propto k\sp{4\dps+1}\ ,\eqn\ener$$
while the enstrophy density is $H(k)=k\sp2E(k)$. Here $\epsilon(k)$ is
the energy density in ${\bf k}$-space.

To choose an appropriate solution from the wealth of conformal solutions,
one should impose some additional conditions that follow from the
symmetries or conservation laws specific for the problem in question.
According to Fjortoft's theorem (see e.g. \lbrack7\rbrack),
the vorticity is the relevant quantity in the problem of small-scale
turbulence (while the energy flux determines large-scale turbulence).
Following Kraichnan \lbrack1\rbrack\ who developed a simple and efficient
(though uncontrollable, of course) closure in terms of double correlation
functions, the enstrophy
$$H_2=\int\omega\sp2({\bf x})d\sp2x\ ,\eqn\Enstrphy$$
which is a motion integral of Euler's equation is usually taken into
account.
A steady turbulence spectrum in the small-scale region should provide for
a constant enstrophy flux over the scales which yields \lbrack5\rbrack\
$$\left\langle{\partial\omega(x+r)\over\partial t}\omega(x)\right\rangle
\propto r\sp0={\rm const}\ .\eqn\flux$$
Puting $\omega=\Delta\psi$ and
$${\partial\omega\over\partial t}\propto \bigl(
L_{-2}\bar L_{-1}\sp2-\bar L_{-2}L_{-1}\sp2\bigr)\phi$$
for the time derivative,
Polyakov obtained \lbrack5\rbrack\
$$(\dph+2)+(\dps+1)=0\ .\eqn\fluxdim$$
As one can see, the enstrophy flux is expressed through the triple
correlation function which can be expressed by the fusion
rule (4) in terms of the double correlation function.

Equation (8) can be obtained also by requiring the rate of the enstrophy
dissipation to remain constant while the viscosity $\nu$ goes to zero:
$${dH_2\over dt}=\nu \int\sp{1/a}k\sp2H_kdk\propto\nu a\sp{-6-4
\dps}\propto \nu\sp{(3+\dph+\dps)/(\dph-
\dps)}\ .\eqn\dissip$$
The last estimate was given by using the expression for the
viscous scale $\nu\propto a\sp{2\dps-2\dph}$
that follows from the comparison of the nonlinear and the viscous terms
in the Navier-Stokes equation.

Kraichnan's dimensional approach would correspond to additive dimensions
$\dph=2\dps$ giving thus $\dps=-1$ and $E(k)\propto
k\sp{-3}$. However, we have arguments that suggest that this in not a
conformal solution from the set of minimal models (see Appendix).

To ensure that the conformal set of correlators is a steady solution,
Polyakov imposed an extra condition requiring
that rhs of (4) (which determines the time derivatives of the
correlators) vanishes in the ultraviolet limit. Since
$$\psi(z_1)\psi(z_2)=(z_1-z_2)\sp{\dph-2\dps}
\phi(z_2)+...$$ as $z_1\rightarrow z_2$, the following inequality arises:
$$\dph>2\dps\ ,\eqn\regul$$ thus, using (8), $\dps<-1$.

Conditions \fluxdim\ and \regul\ by no means determine a single solution.
 The minimal model (2,21) presented by Polyakov \lbrack5\rbrack\ is
nothing but the first example from an infinite family. The curious reader
can find the first few hundred solutions in the Appendix.
Polyakov's solution corresponds to the minimal number of primary
fields (in this case 10). One could, in particular, find the minimal
model (5,72) that gives $\dps=\Delta_{(1,25)}=-7/6$ and
$E(k)\propto k\sp{-11/3}$ as in Moffatt's spectrum.

Usually, when speaking about a turbulent solution carrying a constant
flux, one should check that two conditions are satisfied: i) The solution
should be local in $k$-space which means that distant scales do not
interact substantially; this should be provided by the convergence of the
integral determining the flux in $k$-space;
ii) The constant that arise
in this (converging) integral should be nonzero and have the correct sign
to satisfy the boundary conditions in $k$-space, i.e.,
the pumping and damping. Any solution of (8) and (10) violates both of
these conditions. Inequality (10) means that
$\dps<-1$ i.e. the spectrum (5) is steeper than
Kraichnan's one (which yields an infrared logarithmic divergence) so that
a power infrared divergence arises for any solution. (The correct power
of the divergence can be obtained by using Lagrangian or quasi-Lagrangian
variables which eliminate sweeping of small scales by larger ones
\lbrack8,9\rbrack). For minimal models the three-point function $\langle
\psi\psi\psi\rangle\sim\langle\phi\psi\rangle$ is equal to
zero since primary fields with different dimensions ($\dps
\not=\dph$) are orthogonal. The cases where $\psi$ appears in the
operator product expansion of $\lbrack\psi\rbrack\lbrack\psi\rbrack$
(like those with $\dps=\dph=-3/2$ which one can find in the Appendix)
break parity and therefore should be excluded.
The conformal solutions in question are thus fluxless. The second
difficulty (zero flux) remedies to some extent the first difficulty
(nonlocality) since one should not require the convergence of the
integrals that are identically zero. Physically this corresponds to the
fact that the spectrum of a system in thermodynamic equilibrium should
not be local, unlike the cascade spectrum one. A fluxless spectrum
corresponds to an equilibrium case.

But, since we are discussing a nonequilibrium situation, there
still remains the question, how the spectrum carries nonzero flux from
the pumping region to the viscous region of scales. Polyakov's suggestion
is that small deviations from the power law due to infrared cut-off
(pumping scale) could provide nonzero values for the enstrophy flux. (It
is unclear whether such a flux is local in $k$-space or it is a
noncascade nonlocal solution.) The physical
correlators are thus assumed to be close to the equilibrium ones in
the inertial interval of scales. This is similar to what happens in
two-dimensional optical turbulence described by the
Nonlinear Schr\"odinger Equation
$$i\Psi_t+\Delta\Psi+T\vert\Psi\vert\sp2=0.$$ For wave turbulence,
the small-scale turbulent
spectrum carrying constant energy flux is as follows:
$$\epsilon(k)\propto k\sp{\alpha-m-d}\ ,$$
where $d$ is the space dimension and $\alpha$ and $m$ are the scaling
indices of the Hamiltonian coefficients (i.e. the frequency and
the four-wave interaction coefficient respectively) \lbrack10\rbrack.
For the NSE,
$\alpha=d=2$ and $m=0$ so that the turbulent spectrum is
$\epsilon(k)=$ const which coincides with
the equilibrium equipartition. This spectrum (which is an exact steady
solution) is fluxless too. Numerical simulations of the NSE show the
nonequilibrium spectrum to be close to $\epsilon\approx$ const,
an analytical attempt to introduce a slow logarithmic factor causes some
doubts \lbrack11\rbrack. The same coincidence of the equilibrium and
turbulent spectra takes place for common turbulence of Langmuir
and ion sound waves in plasma, the spectra carrying fluxes acquire
logarithmic factors in this case \lbrack10,12\rbrack.
Polyakov suggested to distort the spectra by analytical (in
${\bf x}$-space) contributions. In our opinion, the main difference of
his approach to hydrodynamic turbulence from the above picture of
wave turbulence, namely,
the assumption that the turbulent spectrum of 2d hydrodynamics should
be close to an equilibrium one, is not based on solid ground.
One would like to see the degeneracy that prescribes the coincidence
of the turbulence spectrum with an equilibrium one.

And what may be more important, Polyakov's considerations do not
take into account the presence of an infinite set of motion integrals
$$H_n=\int\omega\sp{n}({\bf x})d\sp2x\ .\eqn\integ$$
The conservation of $H_n$ follows directly from the fact that the
equation
$${\partial\omega\over\partial t}-e_{\alpha\beta}{\partial\omega\over
\partial x_\alpha}{\partial\over\partial x_\beta}{\delta{\cal H}\over
\delta\omega}=0\eqn\Ham$$
conserves the integral $\int F(\omega)\,dxdy$, where $F$ is an arbitrary
function and the Hamiltonian ${\cal H}$ is an arbitrary functional of
$\omega$ \lbrack not only ${\cal H}=\int\psi\omega\,dxdy$, which gives
the lhs of (1)\rbrack. As one can see, even an infinite number of motion
integrals does not fix the system but only its class.

Most authors feign indifference to the existence
of the infinite number of motion integrals in 2d turbulence.
Some (weak) arguments that only quadratic integrals
(i.e. energy and squared vorticity) should be taken into account
while considering thermodynamic equilibrium
were given by Kraichnan \lbrack\Kraa,\Krad\rbrack.
However, an arbitrary turbulent pump generally produces a nonzero input
of all integrals $H_n$. The theory should describe the fate of these
integrals. Estimating the rate of the viscous dissipation of $H_n$ ($n
\geq2$) similarly to (9), one gets
$${dH_n\over dt}=\nu \int\sp{1/a}k\sp2H_ndk\propto
\nu\sp{\lbrack(n-1)(\dps+1)+\dph+2\rbrack/(\dph-\dps)}\ .\eqn\disint$$
We rewrite it by the help of (8) as follows:
$${dH_n\over dt}=\nu \int\sp{1/a}k\sp2H_ndk\propto
\nu\sp{(n-2)(\dps+1)/(\dph-\dps)}\ .\eqn\disdim$$
Formula (14) again demonstrates that $\dps\leq-1$ while $\dph-\dps<0$,
otherwise the dissipation rate goes to infinity
while viscosity goes to zero. For Polyakov's solution with $\dps<-1$, the
integrals $H_n$ are not dissipated in the inviscid limit when $n>2$.
What is the fate of these integrals if they are injected?
Note that it is impossible to have a local cascade of $H_2$ and
a nonlocal transfer of other vorticity integrals.

One can require a constant dissipation rate of any vorticity integral.
According to \disint\ and \disdim\ it means that $\dps=-1$ and we are
coming back to Kraichnan's spectrum. Indeed, this
corresponds to $\psi(r)\propto \vert r\vert\sp 2$
so that the vorticity $\omega$ has zero scaling index (maybe
logarithmic). All powers of the vorticity can thus have constant fluxes
in $k$-space simultaneously. Actually, Kraichnan's spectrum equally
satisfies all conservation laws.

Returning to the above set of conformal solutions, one can find
subsequences (see Appendix) that give $\dps$ which approaches the value
$-1$, while the number of primary fields in the model increases.
The closer $\dps$ is to $-1$, the better the given conformal
solution satisfies the vorticity conservation laws.
One can suggest to get Kraichnan's spectrum as a limit of such
subsequences, thus considering it as some weak solution. In
hydrodynamics, it is physically intuitive, to account for an infinite
rather than a finite number of primary fields (of topologically different
field configurations). Still, some other
spectra (including those of Saffman and Moffatt) could be realized
for special initial conditions or as intermediate time asymptotics.
Kraichnan's spectrum seems to correspond to the most general conditions
(both initial conditions in time and boundary conditions in $k$-space).
This spectrum seems to be the most ergodic (e.g. isovorticity
lines have the fractal dimension $2$ so that they can fill the entire
space \lbrack13\rbrack). It is quite natural that Kraichnan's spectrum
corresponds to an infinite number of primary fields. How this
spectrum can be distorted by a slow factor to avoid the logarithmic
divergence is, in our opinion, still an open problem.

\ack{Discussions with P.Wiegmann, A.Finkelstein, G.Schutz, R.Plesser,
A.Schwimmer and Y.Levinson are gratefully acknowledged.}
\appendix
In this appendix we will summarize all the conditions set by Polyakov
on the solutions of a turbulence problem. Subsequently we apply these
conditions to the set of minimal models and present a list of a few
hundred solutions. We describe the algorithm we used to compute
these solutions and give arguments why we think $\dps=-1$ is
not a minimal model solution. Finally we present a few solutions
in a sequence that tends to $\dps=-1$.

The conditions are:
\pointbegin $\dps+\dph=-3$
\point $\dps<-1$
\point $\dph$ must be the operator with the smallest dimension in
the OPE of \lbrack$\psi$\rbrack\ and \lbrack$\psi$\rbrack.

The minimal models are characterized by two positive co-prime integers
$(p,q)$. For each minimal model $(p,q)$ there is a set of ${(p-1)(q-1)
\over 2}$ primary fields parameterized by two integers $(n,m)$, $1\leq
n\leq p-1$, $1\leq m\leq q-1$. The spectrum of conformal dimensions is
given by
$$\Delta_{(n,m)}={(nq-mp)^2-(q-p)^2\over 4pq}.\eqn\dnm$$
{}From this formula we see that $\Delta_{(n,m)}=\Delta_{(p-n,q-m)}$ and
that they both correspond to the same primary field. The scaling
dimension of the primary field $(n,m)$ is $2\Delta_{(n,m)}$.
{}From the first condition, using \dnm,
we get ${p\over q}$ in terms of $\nps,\mps,\nph,\mph$
$${p\over q}={\nps\mps+\nph\mph-8+k\over \mps^2+\mph^2-2},\eqn\poqa$$
where $k$ is an integer defined by
$$k^2=(\nps\mps+\nph\mph-8)^2-(\nps^2+\nph^2-2)(\mps^2+\mph^2-2).\eqn\ks
$$ This defines both $p$ and $q$ since they are co-prime.
{}From the third condition and the selection rules for the OPE we get that
$\nph,\mph$ are odd numbers and satisfy $1\leq\nph\leq2\nps-1$,
$1\leq\mph\leq2\mps-1$. From the demand that $\dph$ be the lowest
possible we get $(\nph q-\mph p)\approx0$, this sets
$\mph$ to be the nearest odd to
$$\nph{\nps\mps-8\pm\sqrt{2\mps^2-16\nps\mps+2\mps^2+60}\over\nps^2-2}.
\eqn\mpha$$
An algorithm to calculate these turbulent solutions can be as follows.
Given $\nps,\mps$, one can calculate $\mph$ from \mpha\ and if $k$ (eq.
\ks\ ) is an integer one can olso find $p$ and $q$. Using this algorithm
we got a set of solutions which are listed in the tables below. (This is
a partial list, there is an infinite number of solutions.)

We also want to check if there exists a minimal model for which $\dps=-1$
or $\dph=-2$. We set $\vert\nph q-\mph p\vert$ to its minimal value and
since $p,q$ are co-prime there exist $\nph,\mph$ such that $\nph q-\mph
p=1$ and we get, using \dnm, $-2={1-(p-q)^2\over4pq}$ or $$p^2-10pq+q^2-1
=0.\eqn\pq$$ for large $p,q$ one can neglect 1 and solve the quadratic
equation to get $$\left({q\over p}\right)_\pm=5\pm\sqrt{24},\eqn\qop$$
the two solutions correspond to
the symmetry between $p$ and $q$. (This solution is irrational so in
practice one can take $p,q$ co-prime such that $p\over q$ is
close to this value, one can get any accuracy by taking large $p,q$).
{}From $\dps=-1$ and \qop\ we have $\mps$ in terms of $\nps$ as
the nearest integer to
$$\mps=\nps{q\over p}+2\sqrt{q\over p}.\eqn\mn$$
The sequence defined by
$$a_{k+1}=10a_k-a_{k-1}\qquad a_0=0,\quad a_1=1\eqn\sequence$$
serves as a solution to \pq\ taking $p=a_k$, $q=a_{k+1}$ for any
$k=2,3,...$ the $k$-th term is given by
$$a_k={(5+\sqrt{24})^k-(5-\sqrt{24})^k\over2\sqrt{24}}\eqn\ak$$
Demanding that this $(p,q)$ model has $\dps=-1$ leads to the condition
that $\sqrt{4pq+1}$ is an integer or, in terms of this sequence, that
$${(5+\sqrt{24})^{2k-1}+(5-\sqrt{24})^{2k-1}+14\over24}\eqn\seqsqr$$
is an integer squared, which is unlikely.

In the list below one can find the models with $(p,q)$ close to \qop\
and $(\nps,\mps)$ satisfying \mn:
$(13,129)$, $(39,389)$, $(109,1082)$, $(232,2295)$ and $(69,686)$
with $-\dps={561\over559},{391\over389},{59001\over58969},{17873\over
17748},{7905\over7889}$ respectively.
\bigskip
\refout
\noncenteredtables
\line{\begintable
(p,q)|\multispan{3}\tstrut\hfil$\psi$\hfil|
\multispan{3}\tstrut\hfil$\phi$\hfil\crthick
|n&m|$\dps$|n&m|$\dph$\cr
(2,21)|1&4|$-{8\over7}$|1&7|$-{13\over7}$\cr
(3,26)|1&5|$-{17\over13}$|1&9|$-{22\over13}$\cr
(14,115)|1&6|$-{33\over23}$|1&9|$-{36\over23}$\cr
(14,111)|1&8|$-{56\over37}$|1&7|$-{55\over37}$\cr
(22,179)|1&10|$-{261\over179}$|1&9|$-{276\over179}$\cr
(3,25)|1&11|$-{7\over5}$|1&9|$-{8\over5}$\cr
(26,223)|1&12|$-{297\over223}$|1&9|$-{372\over223}$\cr
(7,62)|1&13|$-{39\over31}$|1&9|$-{54\over31}$\cr
(6,55)|1&14|$-{13\over11}$|1&9|$-{20\over11}$\cr
(34,335)|1&16|$-{69\over67}$|1&9|$-{132\over67}$\cr
(3,25)|2&14|$-{7\over5}$|1&9|$-{8\over5}$\cr
(3,26)|2&21|$-{17\over13}$|1&9|$-{22\over13}$\cr
(11,91)|2&14|$-{108\over77}$|3&25|$-{123\over77}$\cr
(11,87)|2&16|$-{481\over319}$|3&23|$-{476\over319}$\cr
(11,93)|2&20|$-{464\over341}$|3&25|$-{559\over341}$\cr
(8,67)|3&28|$-{369\over268}$|3&25|$-{435\over268}$\cr
(15,119)|4&32|$-{180\over119}$|1&7|$-{177\over119}$\cr
(39,310)|4&31|$-{604\over403}$|5&39|$-{605\over403}$\cr
(13,105)|4&34|$-{19\over13}$|5&41|$-{20\over13}$\cr
(39,361)|4&42|$-{5504\over4693}$|5&47|$-{8575\over4693}$\cr
(13,129)|4&46|$-{561\over559}$|5&49|$-{1116\over559}$\cr
(21,166)|4&31|$-{869\over581}$|7&55|$-{874\over581}$\cr
(9,71)|4&32|$-{319\over213}$|7&55|$-{320\over213}$\cr
(21,172)|4&35|$-{429\over301}$|7&57|$-{474\over14448}$
\endtable\hfil\begintable
(p,q)|\multispan{3}\tstrut\hfil$\psi$\hfil|
\multispan{3}\tstrut\hfil$\phi$\hfil\crthick
|n&m|$\dps$|n&m|$\dph$\cr
(12,97)|5&42|$-{143\over97}$|1&9|$-{148\over97}$\cr
(16,159)|5&56|$-{427\over424}$|3&29|$-{845\over424}$\cr
(12,95)|5&39|$-{3\over2}$|5&39|$-{3\over2}$\cr
(9,71)|5&39|$-{319\over213}$|7&55|$-{320\over213}$\cr
(36,317)|5&48|$-{1219\over951}$|7&61|$-{1634\over951}$\cr
(26,205)|5&39|$-{798\over533}$|9&71|$-{801\over533}$\cr
(13,113)|5&47|$-{1971\over1469}$|9&77|$-{2436\over1469}$\cr
(35,313)|6&58|$-{387\over313}$|1&9|$-{552\over313}$\cr
(43,342)|6&47|$-{3685\over2451}$|3&23|$-{3668\over2451}$\cr
(43,347)|6&50|$-{21948\over14921}$|3&25|$-{22815\over14921}$\cr
(43,422)|6&65|$-{9309\over9073}$|3&29|$-{17910\over9073}$\cr
(59,466)|6&47|$-{20640\over13747}$|5&39|$-{20601\over13747}$\cr
(59,487)|6&52|$-{40467\over28733}$|5&41|$-{45732\over28733}$\cr
(83,654)|6&47|$-{13563\over9047}$|7&55|$-{13578\over9047}$\cr
(115,906)|6&47|$-{5206\over3473}$|9&71|$-{5213\over3473}$\cr
(23,217)|6&62|$-{795\over713}$|9&85|$-{1344\over713}$\cr
(155,1222)|6&47|$-{28407\over18941}$|11&87|$-{28416\over18941}$\cr
(155,1351)|6&56|$-{55047\over41881}$|11&95|$-{70596\over41881}$\cr
(14,111)|7&55|$-{390\over259}$|3&23|$-{387\over259}$\cr
(14,111)|7&56|$-{390\over259}$|3&23|$-{387\over259}$\cr
(14,121)|7&64|$-{1131\over847}$|3&25|$-{1410\over847}$\cr
(12,95)|7&56|$-{3\over2}$|5&39|$-{3\over2}$\cr
(18,175)|7&74|$-{22\over21}$|5&49|$-{41\over21}$\cr
(48,425)|7&66|$-{437\over340}$|7&61|$-{583\over340}$
\endtable}\endpage
\line{\begintable
(p,q)|\multispan{3}\tstrut\hfil$\psi$\hfil|
\multispan{3}\tstrut\hfil$\phi$\hfil\crthick
|n&m|$\dps$|n&m|$\dph$\cr
(42,331)|7&55|$-{3478\over2317}$|11&87|$-{3473\over2317}$\cr
(21,188)|7&67|$-{817\over658}$|11&99|$-{1157\over658}$\cr
(27,236)|7&65|$-{689\over531}$|13&113|$-{904\over531}$\cr
(18,167)|7&70|$-{580\over501}$|13&121|$-{923\over501}$\cr
(63,502)|8&63|$-{378\over251}$|1&7|$-{375\over251}$\cr
(63,589)|8&80|$-{671\over589}$|1&9|$-{1096\over589}$\cr
(71,562)|8&63|$-{30069\over19951}$|3&23|$-{29784\over19951}$\cr
(71,583)|8&68|$-{58812\over41393}$|3&25|$-{65367\over41393}$\cr
(87,686)|8&63|$-{14948\over9947}$|5&39|$-{14893\over9947}$\cr
(87,731)|8&70|$-{29681\over21199}$|5&41|$-{33916\over21199}$\cr
(29,246)|8&71|$-{1617\over1189}$|5&43|$-{1950\over1189}$\cr
(29,274)|8&81|$-{4422\over3973}$|5&47|$-{7497\over3973}$\cr
(111,874)|8&63|$-{24257\over16169}$|7&55|$-{24250\over16169}$\cr
(111,931)|8&70|$-{47393\over34447}$|7&59|$-{55948\over34447}$\cr
(143,1126)|8&63|$-{120786\over80509}$|9&71|$-{120741\over80509}$\cr
(13,107)|8&68|$-{2013\over1391}$|9&73|$-{2160\over1391}$\cr
(183,1442)|8&63|$-{66043\over43981}$|11&87|$-{65900\over43981}$\cr
(61,491)|8&66|$-{43824\over29951}$|11&89|$-{46029\over29951}$\cr
(231,1822)|8&63|$-{15064\over10021}$|13&103|$-{14999\over10021}$\cr
(21,169)|8&66|$-{19\over13}$|13&105|$-{20\over13}$\cr
(33,280)|8&71|$-{15\over11}$|13&111|$-{18\over11}$\cr
(231,2027)|8&74|$-{29223\over22297}$|13&113|$-{37668\over22297}$\cr
(287,2266)|8&63|$-{69897\over46453}$|15&119|$-{69462\over46453}$\cr
(287,2515)|8&74|$-{26529\over20623}$|15&131|$-{35340\over20623}$
\endtable\hfil\begintable
(p,q)|\multispan{3}\tstrut\hfil$\psi$\hfil|
\multispan{3}\tstrut\hfil$\phi$\hfil\crthick
|n&m|$\dps$|n&m|$\dph$\cr
(20,159)|9&71|$-{80\over53}$|1&7|$-{79\over53}$\cr
(40,341)|9&80|$-{459\over341}$|1&9|$-{564\over341}$\cr
(11,87)|9&71|$-{481\over319}$|3&23|$-{476\over319}$\cr
(11,91)|9&77|$-{108\over77}$|3&25|$-{123\over77}$\cr
(22,215)|9&94|$-{489\over473}$|3&29|$-{930\over473}$\cr
(26,205)|9&71|$-{801\over533}$|5&39|$-{798\over533}$\cr
(26,213)|9&76|$-{1312\over923}$|5&41|$-{1457\over923}$\cr
(16,135)|9&79|$-{49\over36}$|7&59|$-{59\over36}$\cr
(25,197)|9&71|$-{1479\over985}$|11&87|$-{1476\over985}$\cr
(25,224)|9&85|$-{99\over80}$|11&99|$-{141\over80}$\cr
(20,181)|9&86|$-{441\over362}$|11&99|$-{645\over362}$\cr
(62,489)|9&71|$-{7597\over5053}$|13&103|$-{7562\over5053}$\cr
(19,150)|9&71|$-{143\over95}$|15&119|$-{142\over95}$\cr
(19,167)|9&83|$-{4107\over3173}$|15&131|$-{5412\over3173}$\cr
(23,202)|9&83|$-{2970\over2323}$|17&149|$-{3999\over2323}$\cr
(33,262)|10&79|$-{198\over131}$|1&7|$-{195\over131}$\cr
(33,268)|10&83|$-{98\over67}$|1&9|$-{103\over67}$\cr
(99,953)|10&102|$-{1027\over953}$|1&9|$-{1832\over953}$\cr
(107,846)|10&79|$-{22753\over15087}$|3&23|$-{22508\over15087}$\cr
(107,891)|10&86|$-{44116\over31779}$|3&25|$-{51221\over31779}$\cr
(123,970)|10&79|$-{5976\over3977}$|5&39|$-{5955\over3977}$\cr
(41,381)|10&98|$-{6028\over5207}$|5&47|$-{9593\over5207}$\cr
(49,386)|10&79|$-{14181\over9457}$|7&55|$-{14190\over9457}$\cr
(21,169)|10&82|$-{1740\over1183}$|7&57|$-{1809\over1183}$
\endtable}\endpage
\end
\line{\begintable
(p,q)|\multispan{3}\tstrut\hfil$\psi$\hfil|
\multispan{3}\tstrut\hfil$\phi$\hfil\crthick
|n&m|$\dps$|n&m|$\dph$\cr
(147,1247)|10&88|$-{2853\over2107}$|7&59|$-{3468\over2107}$\cr
(179,1410)|10&79|$-{12614\over8413}$|9&71|$-{12625\over8413}$\cr
(179,1519)|10&88|$-{369939\over271901}$|9&77|$-{445764\over271901}$\cr
(219,1726)|10&79|$-{94557\over62999}$|11&87|$-{94440\over62999}$\cr
(73,599)|10&84|$-{64128\over43727}$|11&89|$-{67053\over43727}$\cr
(89,702)|10&79|$-{1204\over801}$|13&103|$-{1199\over801}$\cr
(89,748)|10&87|$-{4149\over3026}$|13&109|$-{4929\over3026}$\cr
(19,150)|10&79|$-{143\over95}$|15&119|$-{142\over95}$\cr
(323,2847)|10&92|$-{30885\over23579}$|15&131|$-{39852\over23579}$\cr
(129,1055)|10&84|$-{12928\over9073}$|17&139|$-{14291\over9073}$\cr
(387,3407)|10&92|$-{187979\over146501}$|17&149|$-{251524\over146501}$\cr
(43,396)|10&97|$-{3337\over2838}$|17&157|$-{5177\over2838}$\cr
(459,3670)|10&81|$-{28003\over18717}$|19&151|$-{28148\over18717}$\cr
(27,229)|10&88|$-{928\over687}$|19&161|$-{1133\over687}$\cr
(459,4039)|10&92|$-{37453\over29427}$|19&167|$-{50828\over29427}$\cr